# Testing the Stationarity Assumption in Software Effort Estimation Datasets

Michael Franklin Bosu [1], Stephen G. MacDonell [2], Peter Whigham [3]
[1]*Centre for Information Technology, Waikato Institute of Technology*
[2,3]*Department of Information Science, University of Otago, New Zealand*
[2]stephen.macdonell@otago.ac.nz, [3]peter.whigham@otago.ac.nz

**Abstract**

*Software effort estimation (SEE) models are typically developed based on an underlying assumption that all data points are equally relevant to the prediction of effort for future projects. The dynamic nature of several aspects of the software engineering process could mean that this assumption does not hold in at least some cases. This study employs three kernel estimator functions to test the stationarity assumption in three software engineering datasets that have been used in the construction of software effort estimation models. The kernel estimators are used in the generation of non-uniform weights which are subsequently employed in weighted linear regression modeling. Prediction errors are compared to those obtained from uniform models. Our results indicate that, for datasets that exhibit underlying non-stationary processes, uniform models are more accurate than non-uniform models. In contrast, the accuracy of uniform and non-uniform models for datasets that exhibited stationary processes was essentially equivalent. The results of our study also confirm prior findings that the accuracy of effort estimation models is independent of the type of kernel estimator function used in model development.*

## 1. INTRODUCTION

Software engineering datasets emanate from a complex and dynamic ecosystem that involves numerous actions and interactions of people and technologies over time. Data collected about software projects are used to support decision making during software development and the planning of future projects. This paper focuses specifically on software development effort data that may be used in the ongoing management of the cost and/or schedule of current projects as well as in the estimation of the effort required in future projects. One such aspect is project timing – that is, when in time a project and its constituent activities were undertaken. In ignoring the timing of projects most current effort estimation practices implicitly assume the underlying development processes to be stationary over time. The adoption of the stationarity assumption in SEE has culminated in the treatment of all past data as equally relevant during the modeling process. The key objective of this paper is to test the validity of this stationarity assumption in the context of SEE.

The range of factors that can affect the effort required in software development is vast such as the competence and experience of the developers, the participation of the customer, the commitment of top management, requirements ambiguity, adequacy of tools support and communication among the development team. The list of potential influences is practically endless as demonstrated by the following studies.

Ten factors that have significant influence on the development cost and productivity of software projects were identified when 50 projects were analyzed in a Swedish bank [1]. Wagner and Ruhe [2] divided software productivity factors into two groups; soft factors are deemed to be attributes that are influential over the way people work and technical factors relate to the software itself. Maxwell and Forselius [3] assessed the productivity factors of 206 software projects from twenty-six Finnish companies and found the company and the type of business of the client organization as being the most influential factors.

A potentially important additional aspect missing from the above analyses is that which is in focus here – that is, the *stationarity* of the development process. It is the contention of this study that over some (unknown) period of time, an organization's software development processes will not remain static. In this paper we therefore assess three software effort estimation datasets to determine whether or not their underlying processes remain stationary over time. The rest of the paper is presented as follows. In Section 2 we consider related work. Section 3 describes our research design. Our analysis and results are presented in Section 4, and in Section 5 is the threat to validity of the study. Section 6 is the discussion and conclusion.

## 2. RELATED WORK

Although numerous SEE models have been proposed (see [4]) the number of studies that have considered project timing information in effort estimation is negligible. This section summarizes the few studies that are directly related to the research reported here.

MacDonell and Shepperd [5] assessed the efficacy of two time-aware estimation methods – sequential accumulation of projects over time and a constant moving window of size five – when applied to a proprietary dataset. They obtained improved results over project managers' effort estimates, especially for the moving window approach [5].

Geographically Weighted Regression (GWR) is a method to manage non-stationarity in spatial data. GWR was applied to capture the non-stationarity of relationship in a landscape fragmentation study [6]. GWR derives non-uniform estimates in spatial data; that is, relationships are established in data that belong to a specified (non-uniform) area, as opposed to ordinary least squares regression (OLS) which outputs the estimates of the average or uniform relationships among all observed data.

GWR relies on the assumption that entities that are near to each other in a geographical area are more likely to exhibit similar properties than those that are more distant. This assumption is acted on by weighting nearer areas more than distant areas.

The study here employs a procedure similar to GWR wherein non-uniform weightings are applied to software effort estimation data over time. The use of kernel bandwidth values also enables the determination of the stationarity of the process underlying the data, except that instead of being applied to parameters of space, the approach is applied to the parameters of software projects.

In spite of the proposals of numerous estimation techniques, process (non-)stationarity and its effect on SEE has received minimal attention as reported by Smartt and Ferreira [7]. To the best of our knowledge there are just three prior studies [8], [9], [10] in the software effort estimation domain that have employed kernel estimators in a manner similar to that reported in this paper.

The study presented here differs from that reported by Kocaguneli, Menzies and Keung [9] in that a wider range of kernel bandwidth values (between 1 and 100) is used in order to discover the stationarity properties of the datasets, whereas five selected kernel bandwidth values were used in [9]. In addition, this study employs weighted linear regression to build models based on the sequential accumulation of projects according to their completion dates, while [9] used analogy- based estimation and did not address data accumulation over time. The work presented in this paper has greater similarities with that of Amasaki and Lokan [8] in that it applies linear regression to a growing portfolio of projects using the same set of kernel functions; however, it differs in the use of a wider range of kernel bandwidth values, as they are being applied in this study to assess the stationarity of the datasets, and the processes underpinning the data. The study reported here also employs three datasets exhibiting different characteristics whereas [8] used an extract from the ISBSG repository. Angelis and Stamelos [10] also employed the kernel estimator in software effort estimation based on analogies. They used the kernel function in order to identify the distributions of effort estimates that are not obvious (such as Normal or Lognormal). They [10] used a fixed bandwidth whilst this study uses a range of bandwidths.

The following specific research questions are addressed by this study:

RQ1. Is there only a stationarity process underlying software effort estimation datasets?

RQ2. Does non-stationarity of software effort estimation datasets affect the accuracy of effort estimation models when applied over time?

RQ3. Does kernel type affect the accuracy of software effort estimation models?

## 3. RESEARCH DESIGN

In this section we first describe each of the three datasets to be analyzed along with the particular computation of effort estimation used in each case. We then describe our model development and evaluation process before specifying how the various kernel weightings are determined.

### 3.1 Dataset Descriptions

**NASA93 Dataset**

The NASA93 dataset was collected by NASA from five of its development centers and it collectively represents fourteen different application types. The entire dataset comprises 93 projects undertaken between 1971 and 1987. Projects were completed in the years indicated in the version of the dataset that is available from the PROMISE Repository http://openscience.us/repo/. The dataset is structured according to the Constructive Cost Model (COCOMO81) format developed by Boehm [9]. It comprises 24 attributes of which 15 are the mandatory effort multipliers. Effort multipliers and development modes are describe in detail in [9]. Effort multipliers are assigned a range of predefined values which were obtained from regression analysis of the original COCOMO81 data. The other attributes of relevance are product size, measured in thousands of lines of code (KLOC), and effort, measured in calendar months (where one calendar month is said to be equivalent to 152 person-hours of effort). The computation of effort for COCOMO81 projects is given by equation (1).

$$effort(personmonths) = a * (KLOC^b) * (\prod_i EM_i) \quad (1),$$

where KLOC is size measured in thousands of lines of code and EM represents the effort multipliers. COCOMO81 projects are classified into three development modes that each requires the use of certain parameter values in the model the values of $a$ and $b$ are domain-specific values dependent on the mode of the project being developed.

**Desharnais Dataset**

The Desharnais dataset was collected from ten organizations in Canada by Jean-Marc Desharnais. The projects in this dataset were undertaken between 1983 and 1988. The dataset consists of 81 records and twelve attributes, including size measured in function points and effort measured in person-hours. In most studies that employ this dataset, 77 of the 81 records are used because of missing data in four records [11]. In this study, the version with the 77 projects is therefore also used. The Desharnais dataset, like the NASA93 dataset, contains only the year of project completion and, as such, the training and test data sets are formed in the same way as the NASA93 dataset (i.e., by using the year of project completion).

Though there are twelve attributes in the Desharnais dataset, analysis carried out by Desharnais identified the size and language attributes as those that are influential in a regression model. Kitchenham and Mendes [12] supported Desharnais' claim by proposing the use of the language attribute as a dummy variable. This approach has therefore been adopted in this study for the models developed for this dataset, as shown in equation (2).

$$\ln(effort) = \ln(size) + language \quad (2)$$

This study used the adjusted function points value as the most complete size attribute (rather than the raw function point count) and treated the three-value language attribute as a dummy variable, with the reference dummy value (being the Basic Cobol projects) indicated as "1" in the Desharnais dataset.

**Kitchenham Dataset**
The Kitchenham dataset [13] was collected from the American- based multinational company Computer Sciences Corporation

(CSC). This dataset contains information about 145 software development and maintenance projects that CSC undertook for several clients. There are 10 attributes considered, the size attribute was measured in function points, and effort was measured in person-hours. The attributes also include start date and estimated completion dates, and the projects were undertaken between 1994 and 1999. The attributes useful for effort modeling (based on prior research evidence) are the size attribute and the application type attribute. This study used the application type attribute as a dummy variable with the reference value being the "Development" type. Again following prior work this study uses 105 records related to projects developed for so-called 'client 2' [13].

As this dataset includes information about the actual start date of projects and their duration in days, these values are used to compute each project's completion date. Training sets are formed based on the years in which projects were completed, as was done for the NASA93 and Desharnais datasets. Composition of the test data sets follows a slightly different process, however, because of the availability of actual start dates: a test set consists of projects completed in the subsequent year and started after the date the last project in the training set was completed. This dataset consists of 67 perfective maintenance projects and 38 development projects. The model formulation is shown as equation (3).

$$\ln(effort) = \ln(size) + type \quad (3)$$

**3.2 Effort Estimation Model Development**

In software effort estimation modeling, as in many other fields, the (secondary) dataset is usually split into two, forming a larger training set and a smaller test set. Models are then built using the training set, and the unbiased performance of the models is evaluated on the test set. This study follows a similar approach; the specifics of how the training and test sets are formed are described in the modeling algorithm subsection below. All models in this study are developed using the statistical package R (version 3.5.2). In preparatory testing the Shapiro-Wilk test of normality was applied to the numeric variables in the training sets. All such variables that failed the normality test were logarithm transformed, meaning that in the associated models developed, log(effort) (shown as ln(effort) in the equations) would be the dependent variable and log(size) (ln(size)) one of the explanatory variables. The estimated (natural log) effort values are back-transformed to unscaled values prior to the computation of any accuracy measures. All models are developed using linear regression, considered to be a widely used modeling approach in effort estimation [4]. The actual linear regression equations for each dataset have been presented in subsection 3.1. It should also be noted that the models developed in this study are all *well-formed models*, that is, the degrees of freedom are considered whereby a training set is formed only when the number of projects is at least two plus the number of explanatory variables being used for model construction.

**Modeling Algorithm**
This paper generally follows the sequential accumulation approach used by MacDonell and Shepperd [5] in forming the training sets for the effort estimation models. As such, the following procedures are applied to all datasets modelled in this study:

1. For each dataset with timing information, select the first year in which projects were completed as the training set – if the first year of projects comprises fewer than the number of observations needed to build a well-formed model, add the next year(s) of projects until the minimum requirement for a well- formed model is satisfied. The subsequent year of projects is used as the test set.
2. Check for normality in the distributions of the training data – if data follow a normal distribution go to step 3 else
2.1. Apply the appropriate transformation to make the data normal and recheck normality for verification as above.
3. Build a regression model using the training data.
4. Apply the model obtained in step 3 to predict the values in the test set.
5. Calculate the accuracy measures (see below) for the prediction model.
6. Add the test year's data to the training set, and the subsequent year's data becomes the new test set.
7. Repeat steps 2 to 6 through to the estimation of the last year of projects.

**Model Evaluation**
We employ the relative error (RE) measure in evaluating each of the models developed in this study. This is because the relative error measure accounts for the variability in data and as such it is robust to outlier data points [14]. Values of RE equal to or greater than 1 indicate that the model is performing no better than the prediction of a constant value [14], while values approaching zero indicate an increasingly accurate prediction. The relative error is computed using equation (4):

$$RE = variance(residuals)/variance(measured) \quad (4),$$

where measured is test data.

### 3.3 Generation of Kernel Weights

In order to apply a consistent approach to our analysis the completion date of each project in the three datasets is the only property of time considered in the determination of the kernel weights in this section (even though the Kitchenham datasets include project start *and* completion dates).

Table 1. Formulae of Kernel Types

| Kernel Type | Formula |
|---|---|
| Uniform | $W_{ij} = 1$, $|t| < 1$ |
| Gaussian | $W_{ij} = \exp(-0.5 * t^2)$, $|t| < 1$ |
| Epanechnikov | $W_{ij} = 1 - t^2$, $|t| < 1$ |
| Triangular | $W_{ij} = 1 - |t|$, $|t| < 1$ |

Table 1 shows the four kernel estimators used in generating weights applied to the datasets (where the Uniform kernel serves as a non-weighted baseline). To find t, we used formula (5):

$$t_{ij} = (t_j - t_i)/b \quad (5)$$

where $t_{ij}$ is the period, or in this case, the number of years that have elapsed between project i and the target project j (that is, the project being estimated). $W_{ij}$ is the weight applied to project(s) completed in year i with reference to projects in a target year j, and b is the kernel bandwidth (discussed later in this section).

The value of 1 is assigned to the oldest completion period in each dataset and a yearly increment of 1 is applied thereafter.

The elapsed time periods are determined between a specific year and the target year to be used in the application of the formulae in Table 4 to derive the weights for projects in specific years; each past year is subtracted from the target year and the results indicate the elapsed time (in years) from the target year. For instance, given two projects developed in different years; $t_{ij} = j - i$. The weight is 1 when i is equal to j. The band width controls the weighting contribution of neighboring projects, that is, projects from specific years [8].

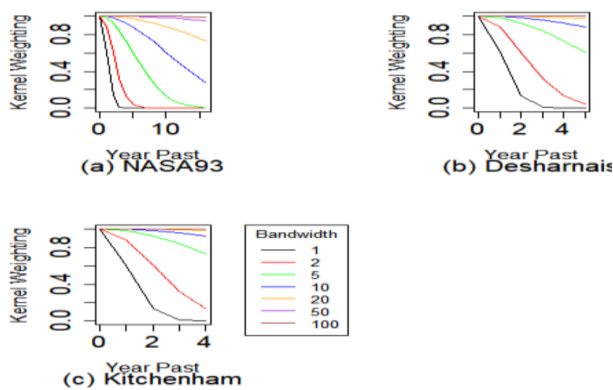

Fig. 1. Weights generated for datasets using the Gaussian Kernel

Fig. 1 depicts the weights that are generated for selected bandwidth values for the datasets based on the Gaussian kernel used in this paper (Note that for clarity, it is impractical to show all the bandwidth values between 1 and 100). For this study, the bandwidths are set between 1 and 100 at increments of 1. Fig. 1 shows that, as the bandwidth value increases, the weights applied to all projects in the training set approach 1. Older projects have smaller weights because the assumption is that the underlying software process used in generating the data is different to that used for current projects. It is also evident in Fig. 1 that small bandwidth values such as 1 and 2 lead to a rapid decline in the weights that are assigned to projects that occur later in time from the target year.

However, the weight for larger bandwidth values declines gradually and as such the weights for the data in the training set become nearly the same irrespective of the completion date of a project. Due to lack of space, all other graphs generated for selected bandwidth values for all the datasets and kernel types are not shown, however, they are available at this link[1]. The concave nature of the Epanechnikov kernel for the NASA93 dataset curves corresponds to the expected shape of this particular kernel [8]. In comparison to the Gaussian kernel curves the weights decrease a little more gradually, for all bandwidth values across the periods of project completion. Finally, the weights generated for selected bandwidth values for the datasets based on the Triangular kernel are linear for all bandwidth values and across all periods. Just like the Gaussian and the Epanechnikov kernels, the weights for larger bandwidth values decline in a more gradual manner.

### 4. ANALYSIS AND RESULTS

The kernel weights generated as per the procedure described in subsection 3.3 are applied to effort estimation models for the three datasets. The relative errors of the models are computed over the specified range of bandwidth. Use of the kernel functions enables the application of non-uniform weights to the projects in these datasets as they are used to develop effort estimation models. In order to determine the stationarity or otherwise of these datasets, effort estimation models are developed according to the modeling algorithm of subsection 3.2. The modeling equations derived for each of the datasets in subsection 3.1 are subsequently applied.

In order to determine whether or not a model exhibits a stationary process, the weight graph (Fig. 1) above should be considered alongside the graphs depicting prediction errors on Fig. 2 and others available at the previously specified link. For example, in the case of the Gaussian kernel, Fig. 1 is read in combination with graphs of the models developed for each of the three datasets that used the Gaussian kernel in weight generation, shown in Fig. 2. The bandwidth at which stationarity was attained is identified on the graph of the respective dataset and then this bandwidth value is mapped onto the corresponding Fig. 1 curve to determine the year at which the models remained stationary. This process is repeated for all kernel types and datasets (available at previously specified link) in the interpretation of the results.

---

[1] https://tinyurl.com/SEKE2020-Stationary-Analysis

The accuracy measure of the models built using the weights generated by the kernel estimators are shown on the plots as 'train', which is effectively the non-uniform model (applying non-uniform weighting). The non-uniform model is then used to predict the effort of projects in the test set, indicated as 'test' on the graphs. Similarly, the result of the uniform model (where no weighting is applied) is indicated on the plot as 'train global', and the model is then used to predict the effort of projects in the test set, indicated as 'test global'. The results are shown on each graph to aid comparison of the models and to enable the identification of models that are stationary or otherwise. It is worth noting that, in presenting the results, emphasis is placed on the training model outcomes because the intention is to identify the stationarity status in the data. The results are subsequently presented for each of the datasets in this section, however, only Gaussian kernel for the NASA93 datasets will be illustrated in detail due to lack of space. The other datasets and kernel types follow the same procedure outlined in section 4.1.

### 4.1 NASA93 Dataset

The results of the models developed for the Gaussian kernel modeling of the NASA93 dataset are shown in Fig. 2. The graphs show the relative error against bandwidth values for models built over the various time periods under consideration. In Fig. 2(a), at approximately bandwidth 5, the non-uniform model and the uniform model converge, meaning a stationary process is achieved at this point. Looking at a bandwidth of 5 on Fig. 1(a) indicates that convergence would occur at about the 15th year of projects in the training set. Given that the training set for this model is made up of only 7 years of projects, this means there is effectively no convergence, implying that these projects exhibit a non-uniform process. The underlying process can therefore be said to be non-stationary. The results of the model depicted in Fig. 2(c) are similar to those shown in Fig. 2(a). These two models, Fig. 2(a) and Fig. 2(c), converge at about a bandwidth value of 5. According to Fig. 1(a), a bandwidth value of 5 converges beyond the number of years that constitute the entire NASA93 dataset, implying that the model of Fig. 2(c) also exhibits a non-stationary process.

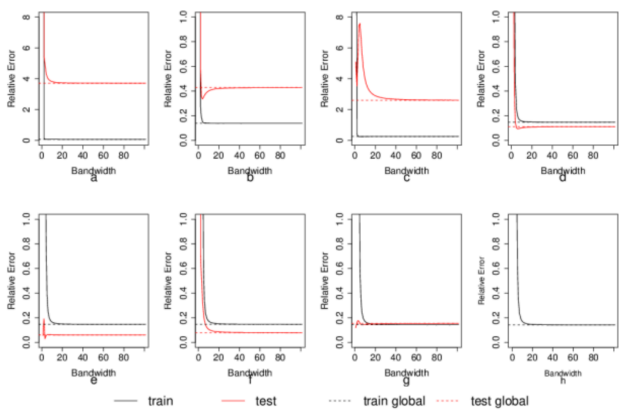

Fig. 2. Gaussian Models - Relative Error against Bandwidth for the NASA93 Dataset.

Fig. 2(d) indicates that at about bandwidth 14 the model started converging and that actual convergence occurred at bandwidth of 25, which according to Fig. 1(a) is well beyond the number of years of projects that constitute the training set, implying that all of the projects that constitute the non-uniform model exhibited a non-stationary process.

The non-uniform model of Fig. 2(e) started approaching a stationary process at a bandwidth value of about 17. If this is mapped onto Fig. 1(a), it is beyond the number of years for which convergence can be attained based on the training set, implying that the model exhibits non-stationary characteristics.

The non-uniform models of Figs. 2(f) and 2(g) both started approaching the curve of the uniform model at a bandwidth value of around 20. The actual convergence of the non-uniform models to the uniform models occurred at bandwidth of 30 and 35 respectively on Fig. 2(f) and Fig. 2(g). This again occurs beyond the number of years of projects in the datasets (as indicated on Fig. 1(a)) which implies that the projects used in building the models exhibited non-stationary characteristics.

A model developed using the entire NASA93 dataset, as shown in Fig. 2(h), started approaching the uniform model curve at bandwidth 15 and actually converged to that of the uniform model at about bandwidth 18. This convergence value according to Fig. 1(a) requires more than the 14 years of projects that constitute the NASA93 dataset, implying that the process underlying this model is non-stationary.

Overall Fig. 2 indicates that the accuracy of the uniform models is better than (that is, they exhibit lower relative error values) the non-uniform models for the NASA93 dataset. The curves also show the existence of non-stationary processes underlying the projects of the NASA93 data set across the different projects over time, evident in the rapid decline of the relative error of the non-uniform models as the bandwidth value increases.

### 4.2 Desharnais Dataset

The results for the Desharnais dataset using the Gaussian kernel function indicate that, in general, the uniform models are nearly the same as the non-uniform models in terms of their accuracy, though the non-uniform models are marginally better in two cases. For the model built with the entire Desharnais dataset, the non-uniform and the uniform model results are nearly the same, with both exhibiting an underlying stationary process. Taken overall, the results of the Desharnais model analysis generally indicate a nearly stationary process across the different bandwidths and across time. For this dataset, the non- uniform model and uniform model predictions are nearly the same, for all the models. The predictions based on the models (non-uniform and uniform) is similar to that described for the NASA93 dataset in section 4.1, as some of the models' predictions are better in terms of accuracy than others across time. The results obtained for Epanechnikov kernel and the Triangular kernel are largely similar to those obtained for the Gaussian models.

The relative stationarity of the models built for the Desharnais dataset is somewhat surprising because this dataset was collected from 10 different organizations in Canada over a period of 7 years. However, the project types and development languages used were few. This perhaps implies that it is possible that organizations working at the same time on similar projects may well have similar practices and, as such, models that are built to characterize their practices may be more homogeneous than heterogeneous.

**4.3 Kitchenham Dataset**
The models developed using the Gaussian kernel when applied to the Kitchenham dataset depicts a near stationary process. The first models exhibits non-stationarity at lower bandwidth values until they converge to a stationary process at bandwidth values of between 4 and 10. Mapping these bandwidth values onto Fig. 1(c) indicates that the models will not attain stationarity in time – thus the process underlying the second to fourth models are interpreted as being non-stationary. The results for this dataset are therefore mixed – there is evidence of a stationary process in one of the models while the other three imply a non-stationary process. The predictions based on the Gaussian model are relatively good for this dataset as they all attained a relative error of less than 1.

The results of the models based on the Epanechnikov and Triangular kernels are similar to their equivalent Gaussian curves based on exact bandwidth values comparisons (they exhibit similar stationarity and non-stationarity at the same bandwidth values respectively). The accuracy of the models of all three kernel estimators are similar for the Kitchenham dataset. Both the respective non-uniform models and the uniform models generated similar results in terms of the RE measure. The predictions based on the models differ across time as was observed for the previous two datasets.

The mixed results (both stationary and non-stationary models) obtained for the Kitchenham dataset could be attributed to the different practices associated with the development types: it seems likely that the organization would have applied different processes to new software development projects as compared to their perfective maintenance projects. This could explain the non-stationarity of some of the models. On the other hand, the stationary model could be due to the fact that all projects were developed by one organization for a single client, and as such, similar general (organization-level) procedures could have been applied.

## 5. THREATS TO VALIDITY
The first threat to the validity of this study is to the generalization of our results, as the datasets used are convenience sampled from the PROMISE repository. Though these datasets cannot be considered as representative of the entire software industry, those stored in the PROMISE repository have rather become benchmark datasets in empirical software engineering. Moreover, the three datasets were selected in terms of their possessing different characteristics. As such these results provide promising insights into the derivation of the nature of processes underlying software engineering datasets, and the effect of stationarity on the effectiveness of non-uniform or uniform estimation models.

Another threat is the lack of detailed information for the publicly available datasets. The absence of data detailing the composition of the development teams, the experience of the team and manager, the tools that supported the software development process, the procedures applied at the different development phases, and so on, means that we characterized models as stationary or non-stationary due to the nature of their curves when plotted. Whether the underlying datasets are truly in keeping with this characterization cannot be determined from the limited data available.

## 6 DISCUSSION AND CONCLUSION
Three kernel estimator functions have been applied to three datasets in developing non-uniform models that identifies the stationarity and/or non-stationarity process underlying SEE datasets. Based on the results presented, the research questions are answered as follows.

**RQ1. Is there only a stationarity process underlying software effort estimation datasets?**
The result of this study indicates that for the datasets used in this study, both stationary and non-stationarity processes might be present in software effort estimation datasets. The result further establishes that, it is even possible for one dataset to exhibits both stationary and non-stationary process over time as evidenced by the Kitchenham datasets.

**RQ2. Does non-stationarity of software effort estimation datasets affect the accuracy of effort estimation models when applied over time?**
In considering the above results we determine that the answer to research question RQ2 is yes. For all datasets that exhibited non-stationary processes the models (non-uniform models) resulted in relatively large relative errors especially prior to convergence to the uniform models. In contrast, the estimation accuracy for datasets that exhibited stationarity is in almost all cases the same as that obtained from the uniform models. These results are observed for all kernel types. Thus, we would conclude that the accuracy of effort estimation models is indeed affected by the stationarity of the datasets.

**RQ3. Does kernel type affect the accuracy of software effort estimation models?**
For the datasets that have been analyzed in this study the evidence indicates that the type of kernel does not affect model accuracy. The accuracy of the models as measured by the relative error were mostly the same for the respective datasets for all kernel types. The estimations based on the models using the test sets were also the same for each dataset irrespective of the kernel type that was used in the generation of the weights. This study therefore reaffirms the result of the Kocaguneli, Menzies and Keung [9] study that did not find variation in model accuracy due the type of kernel. In terms of using different kernel types to assess the stationarity of a dataset there were just a few occurrences where the different kernel types generated contrasting results, as presented in Section 4.

This study found that there is the possibility of both stationarity and non-stationarity processes present in SEE datasets. A further finding is that the stationarity or otherwise of a datasets impacts on model prediction accuracy. The evidence drawn from this study further suggests that the accuracy of models is independent of the kernel type used in the generation of weights for the non-uniform models. This is observed in the fact that, for each dataset, all three kernel estimators resulted in the same relative errors for all equivalent models and their estimates for the test set observations.

Future work will apply the kernel estimators to other datasets as well as assess the effect of bandwidth values on model accuracy.